\documentclass{cernyrep}
\usepackage{wasysym}
\begin{document}
\title{A CERN-based high-intensity high-energy proton source for
long baseline neutrino oscillation experiments with next-generation large underground 
detectors for proton decay searches and neutrino physics and astrophysics\footnote{Based
on a document submitted 
to the CERN SPC Panel on Future Neutrino Facilities (November 2009).}}

\author{Andr\'e Rubbia}
\institute{ETH Zurich, 101 Raemistrasse, CH-8092 Zurich, Switzerland \\and\\
High Energy Accelerator Research Organization (KEK), Tsukuba, Ibaraki 305-0801, Japan}

\maketitle

\begin{abstract}
The feasibility of a European next-generation very massive neutrino observatory
in seven potential candidate sites located 
at distances from CERN ranging from 130~km to 2300~km, is
being considered within the LAGUNA design study.  
The study is providing a coordinated technical design and assessment of
the underground research infrastructure in the various sites, and its 
coherent cost estimation.
It aims at a prioritization of the sites within summer 2010 and a start
of operation around 2020.
In addition to a rich non-accelerator based physics programme including the GUT-scale
with proton decay searches, the detection of a next-generation neutrino superbeam tuned to measure the 
flavor-conversion oscillatory pattern (i.e. 1st and 2nd oscillation maxima) would allow
to complete our understanding of the leptonic mixing matrix, in particular
by determining the neutrino mass hierarchy and by studying CP-violation in the
leptonic sector, thereby addressing the outstanding puzzle of the origin of the excess of matter over 
antimatter created in the very early stages of evolution of the Universe. 
We focus on a multi-MW-power neutrino superbeam (=``hyperbeam'') produced by high-intensity primary protons of energy 30$\div$50~GeV.
We argue that this option is an effective way to establish long baseline neutrino
physics in Europe with the high-stake prospects of measuring $\theta_{13}$ and addressing
CP-violation in the leptonic sector.
\end{abstract}

\section{Physics goals}\label{sec:goals}

Large underground neutrino detectors, like SuperKamiokande~\cite{Kajita:2006gs}
and SNO~\cite{Ahmad:2002jz}, have achieved fundamental results in particle and astro-particle
physics. The construction in Europe of 
next-generation very large multipurpose neutrino observatory
of a total mass in the range of 100'000 to 1'000'000 tons
 devoted to particle and astroparticle physics was recently discussed~\cite{Autiero:2007zj}.
Such a massive detector will provide new and unique scientific opportunities 
in this field, likely leading to fundamental discoveries, and
 is currently listed as one of the priority of the ASPERA roadmap defined in 
 2008~\cite{aspera}.  

 The FP7 Design Study LAGUNA ~\cite{laguna} (Large Apparatus studying Grand Unification and Neutrino Astrophysics)
  is a EC-funded project carrying on underground sites investigations and design for such an observatory.
 Three detector options are currently being studied: GLACIER~\cite{Rubbia:2004tz}, LENA~\cite{Oberauer:2005kw}, 
 and MEMPHYS~\cite{deBellefon:2006vq}.

The new observatory aims at a significant improvement
in the sensitivity to search for proton decays, pursuing the only
possible path to directly test physics at the GUT scale, extending
the proton (and bound neutron) lifetime sensitivities up to $10^{35}$
years, a range compatible with several theoretical models~\cite{Bueno:2007um}; moreover it will
detect neutrinos as messengers from astrophysical objects as well
as from the Early Universe to give us information on processes happening
in the Universe, which cannot be studied otherwise. In particular,
it will sense a large number of neutrinos emitted by exploding galactic
and extragalactic type-II supernovae, allowing an accurate study of
the mechanisms driving the explosion. The neutrino observatory will
also perform precision studies of other astrophysical or terrestrial
sources of neutrinos like solar and atmospheric ones, and search for
new sources of astrophysical neutrinos, like for example the diffuse
neutrino background from relic supernovae or those produced in Dark
Matter (WIMP) annihilation in the centre of the Sun or the Earth. 

Coupled to advanced neutrino beams from CERN, it would
measure with unprecedented sensitivity the last unknown mixing angle
$\theta_{13}$, determine the neutrino mass hierarchy and unveil the existence of CP violation
in the leptonic sector, which in turn could provide an explanation
of the matter-antimatter asymmetry in the Universe.

\section{Main goal of the LAGUNA design study}
Europe currently hosts four national underground laboratories located resp. 
 in Boulby (UK), Canfranc (Spain), Gran Sasso (Italy), and Modane (France),  
 with detectors looking for Dark Matter or neutrino-less double beta decays, or performing long-baseline experiments. 
 However, none of these existing laboratories is large enough for the next-generation very massive
 neutrino experiments.
The LAGUNA design study is therefore evaluating possible extensions of the existing deep underground laboratories, 
and on top of it, the creation of new laboratories in the following regions: Umbria Region (Italy), Pyh\"asalmi (Finland), 
Sierozsowice (Poland) and Slanic (Romania). 

Table~\ref{tab:1} 
summarizes some basic characteristics of the sites under consideration. It also lists 
their distance from CERN and the neutrino energies corresponding to the first maximum of the oscillation for 
the present estimate of the mass squared difference 
$\Delta m^2_{23}\sim 2.5\times 10^{-3}$~eV$^2$.
These are relevant to optimize the energy spectrum of the neutrino beam.
In order to consider all possible baselines, the new CERN
neutrino superbeam should provide neutrinos in an energy range $0\div 7$~GeV.
The actual optimization depends of course on the chosen site.

\begin{table}[h]
\begin{center}
\caption{Potential sites being studied with the LAGUNA design study.}
\label{tab:1}
\begin{tabular}{ccccc}
\hline\hline
\textbf{Location} 	& \textbf{Type} & \textbf{Envisaged depth}	 
								& \textbf{Distance from}	
									& \textbf{Energy 1$^{st}$ Osc. Max.}	 	\\
				&	& \textbf{m.w.e.} 	& \textbf{CERN [km]}		& \textbf{[GeV]} 	\\												
\hline
Fr\'ejus (F) 		& Road tunnel  &  $\simeq$ 4800 & 130 & 0.26   \\
Canfranc (ES)  		& Road tunnel  &  $\simeq$ 2100  & 630 & 1.27  \\
Umbria(IT) $^a$ 		& Green field    &  $\simeq$ 1500  & 665  & 1.34  \\
Sierozsowice(PL)  	& Mine 	         &  $\simeq$ 2400  & 950 & 1.92  \\
Boulby (UK)	  	& Mine 		&  $\simeq$ 2800  & 1050 & 2.12  \\
Slanic(RO)	  	& Salt Mine 	&  $\simeq$ 600  & 1570 & 3.18  \\
Pyh\"asalmi (FI)	& Mine 		&  up to $\simeq$ 4000  & 2300 & 4.65  \\
\hline\hline
\multicolumn{3}{l}{$^{a}$ \footnotesize{$\simeq$1.0 $^{\circ}$ CNGS off axis.}}
\end{tabular}
\vspace*{-7mm}
\end{center}
\end{table}


Site selection is a complex process involving the optimization and assessment of several parameters,
encompassing physics performance, technical feasibility, safety and legal aspects, socio-economic and environmental impact,
costs, etc. As a result, LAGUNA is an interdisciplinary study, involving most European physicists interested in the physics
of massive underground neutrino detectors, as well as 
 geo-technical experts, geo-physicists, structural engineers, mining engineers and also large storage tank engineers.
It  regroups 21 beneficiaries, composed of academic institutions from Denmark, Finland, France, Germany, Poland,
Spain, Switzerland, United Kingdom, as well as industrial partners specialized in civil and mechanical engineering
and rock mechanics, commonly assessing the feasibility of a this Research Infrastructure in Europe.

The study, which started during the summer 2008, is well advanced and
interim reports for each of the seven sites are being compiled\footnote{The present versions of the
interim reports are in the range of 100-200~pages per document. The public documents are available upon request.}. Several documents
will be published within the end of the study. The consortium intends to converge
to a prioritized list of sites within the summer 2010.

\section{Frontier technologies for next generation neutrino detectors}
The search for proton decays with lifetimes up to $10^{35}$~years
as well as the
measurement of the unknown mixing angle $\theta_{13}$ and the
prospects to discover CP-violation in the leptonic sector and to determine
the neutrino mass hierarchy,
make clear that the next generation neutrino experiments will be more ambitious than previous
ones. 

European groups are actively engaged in the R\&D in technologies based on large volume liquids for 
future neutrino detectors, although this work is not funded as part of the LAGUNA design study. We briefly
describe these activities, subdivided into the three different detector options:
\begin{itemize}
\item {\bf Water Cerenkov Imaging Detector}: MEMPHYS is envisioned as a 0.5~Mton scale
detector extrapolated from the Super-Kamiokande and consisting of 3 separate
tanks of 65~m in diameter and 65~m height each. Such dimensions meet the
requirements of light attenuation length in (pure) water and hydrostatic pressure on the bottom PMTs.
A detector coverage of 30\% can be obtained with about 81'000 PMT of 30~cm diameter per tank.
MEMPHYNO~\cite{memphyno} is one R\&D
item in Europe whose main purpose is to serve as test bench for new photo-detection
and data acquisition solutions. Based on the extensive experience of Super-Kamiokande,
this technology is best suited for single Cerenkov ring events typically occurring at energies below 1~GeV.
\item {\bf Liquid Argon Time Projection Chamber (LAr TPC)}: application of this technology, originally developed
at CERN,  to large
detectors was pioneered in Europe by the ICARUS effort which culminated in the successful
operation of half-T600 on surface~\cite{Amerio:2004ze}. After several years of installation
at LNGS,  underground operation of T600 is expected soon.
GLACIER  is a proposed scalable concept for single volume very large detectors up to 100~kton. 
The cryostat is based on industrial liquefied natural gas (LNG) technology
and ionization imaging readout relies on the novel
LAr LEM-TPC~\cite{Badertscher:2009av}, 
operated in double phase with charge extraction and amplification in the vapor phase.
The corresponding R\&D
programme (see e.g. Refs.~\cite{Rubbia:2009md, Marchionni:2009tj}) is 
taking place at CERN and KEK (e.g. CERN RE18/ArDM~\cite{Rubbia:2005ge} is a small-scale 1~ton detector
of the GLACIER-design, being commissioned at CERN; the LEM/THGEM are developed in 
Collaboration with RD51~\cite{RD51}).
European and Japanese groups are collaborating towards the realization
of very large 100~kton-scale detectors. 
The powerful imaging is expected to offer excellent conditions
to reconstruct with high efficiency electron events in
the GeV range and above, while considerably suppressing
the neutral current background mostly consisting of misidentified $\pi^0$'s.
\item {\bf Non-segmented liquid Scintillator Detectors}: LENA is proposed
as a 50~kton liquid scintillator tank of height 100~m and diameter of 26~m,
surrounded by 2~m of water for vetoing external muons. The scintillation
light produced is detected by 12'000 photomultipliers of 50~cm diameter each.
Intensive R\&D on liquid scintillators and photo-sensors has been carried
out in the last years~\cite{MarrodanUndagoitia:2008zz}.
BOREXINO~\cite{Alimonti:2008gc} is a large liquid scintillator detector presently operating
at Gran Sasso. The capabilities to study and identify neutrino 
beams events is being addressed~\cite{Juhaprivate}.
\end{itemize}
%
A comparison of the  physics performance of the three detector options is planned 
within the WP4 of the LAGUNA design study. These simulations include
the physics reach of long baseline experiments from neutrino beams from CERN.


\section{Physics goals of next generation long baseline experiments}

We can consider the following goals for next generation long baseline experiments beyond the current
round of approved experiments:
\begin{enumerate}
\item Detect of the $\nu_\mu\rightarrow \nu_e$ in the appearance mode and measure $\theta_{13}$ or
improve limit on $\theta_{13}$ by an order of magnitude compared
to current reactors, T2K~\cite{Itow:2001ee} and NOvA~\cite{Ayres:2004js};
\item Measure the CP violation in the leptonic sector parametrized by $\delta_{CP}$;
\item Determine the neutrino mass hierarchy;
\item Detect of the $\nu_\mu\rightarrow \nu_\mu$ in the disappearance mode
to reduce the experimental errors on  $\Delta m^2_{32}$ and $\theta_{23}$ and observe the $\nu_\mu$
oscillation behavior (of the 1st and 2nd maximum);
\item Detect $\nu_\mu\rightarrow \nu_\tau$ in appearance mode with statistics significantly improved
compared to OPERA~\cite{Agafonova:2009kr};
\item Measure the ``solar'' product $\Delta m^2_{21}\times \sin^2\theta_{12}$ in the $\nu_\mu\rightarrow \nu_e$
appearance channel. 
\end{enumerate}
We postpone the discussion of the three last measurements, point 4 being limited by systematic errors. 
We however point out that they are as relevant as the first three.

The discovery of $\theta_{13}$ is determined by the ability to
detect a statistically significant excess of $\nu_e$~charged current events above the
predicted backgrounds, and its sensitivity scales therefore as $S/\sqrt{B}$. 
Hence, the next generation experiments require significantly more
event rate and better background rejection compared to present round T2K and NOvA, and should aim
to improve the $\theta_{13}$ statistical sensitivity by at least one order.
A precise knowledge of the beam flux and neutrino cross-sections is also mandatory, to contain systematic errors
-- e.g. the $\nu_e$ contamination of conventional beam is on the
order of 1\% of $\nu_\mu$, hence a 10\% systematic error is equivalent
to a systematically limited sensitivity of $\sin^22\theta_{13}\approx O(1\permil)$ on the appearance signal.
The study of CP-violation is more challenging
since it requires to measure the oscillation probability as a function of the
neutrino energy, or alternatively to compare large samples of $\nu_e$ and $\bar\nu_e$ 
CC~events, and suffers in general from neutrino oscillation parameters degeneracies.

To illustrate the case, we consider the T2K sensitivity
$\sin^{2}2\theta_{13}>0.01$ (90\%C.L.) obtained with 
22.5 kton fiducial mass of Superkamiokande and 5 years of neutrino running at a 
proton beam power of 750 kW.
If an excess is found in T2K, we can envisage a precise measurement of the $\nu_{\mu}\rightarrow\nu_{e}$
oscillation probability with an increase of beam intensity up
to 1.66 MW ($\simeq \times2$), a partial re-optimization of the flux within the constraints
of an existing beamline infrastructure  -- longer baseline
but smaller off-axis angle to Okinoshima island to increase beam energy  ($\simeq \times1$)
-- and  a 100~kton liquid Argon TPC
($\simeq \times4.5$) but higher detection efficiency ($\simeq \times2$) and lesser
background ($\simeq \div 2$) \cite{Badertscher:2008bp}, for an overall
gain of $(2\times 4.5\times 2)/\sqrt{2\times 4.5/2}\approx 10$.
See Ref.~\cite{Hasegawa:2010xy} for details.

Similar considerations apply to the US scenarios~\cite{Kimyp}.

\subsection{Measurement of the oscillation probability as a function of energy}
The neutrino flavor oscillation probability including atmospheric, solar and interference
terms, as well as matter effects, can expressed using the following 
equation~\cite{Freund:2001pn,Cervera:2000kp,Hagiwara:2006vn}
\begin{equation}
P(\nu_{e} \rightarrow \nu_{\mu}) \sim \sin^{2} 2\theta_{13} \cdot T_{1} + \alpha \cdot \sin \theta_{13} \cdot (T_{2}+T_{3}) + \alpha^{2} \cdot T_{4} .
\end{equation}
where, 
\begin{eqnarray}
T_{1} &=& \sin^2 \theta_{23} \cdot \frac{\sin^2 [(1-A) \cdot \Delta]}{(1-A)^{2}} \nonumber\\
T_{2} &=& \sin \delta_{CP} \cdot \sin 2\theta_{12} \cdot \sin 2\theta_{23} \cdot \sin \Delta \frac{\sin(A\Delta)}{A} \cdot \frac{\sin[(1-A)\Delta]}{(1-A)}  \nonumber\\
T_{3} &=& \cos \delta_{CP} \cdot \sin 2\theta_{12} \cdot \sin 2\theta_{23} \cdot \cos \Delta \frac{\sin(A\Delta)}{A} \cdot \frac{\sin[(1-A)\Delta]}{(1-A)}  \nonumber\\
T_{4} &=& \cos^2 \theta_{23} \cdot \sin^2 2\theta_{12} \frac{\sin^2(A\Delta)}{A^2}.
\end{eqnarray}
where 
$\alpha \equiv \frac{\Delta m_{21}^2}{\Delta m_{31}^2}$, $\Delta \equiv
\frac{\Delta m_{31}^2 L}{4E}$, $A \equiv \frac{2\sqrt{2} G_{F} n_{e}
E}{\Delta m_{31}^2}$. $\Delta m^2_{31} = m_3^2 - m_1^2$, 
$\Delta m^2_{21} = m_2^2 - m_1^2$, $\theta_{13}$ is the mixing angle of
the 1st and 3rd generations, while $\theta_{12}$ is that for 1st and 2nd,  
and $\theta_{23}$ is that for 2nd and 3rd generations. 

From the analysis of this expression, it can be noted that the effects of the CP phase $\delta_{CP}$ appear 
as either~\cite{Badertscher:2008bp}:
\begin{itemize}
\item a difference between $\nu$ and $\bar{\nu}$ behaviors which corresponds to the
sign changes $\delta_{CP}\rightarrow -\delta_{CP}$ and $A\rightarrow -A$
(this method is sensitive to the $CP$-odd term $T_2$ which vanishes
for $\delta_{CP}=0$ or 180$^\circ$);
\item a formal parametric dependence of the energy spectrum shape of the appearance oscillated $\nu_e$ as
given by the formula $P(\nu_{\mu} \rightarrow \nu_e)$ as a function of $\delta_{CP}$  and all other oscillation parameters
(this method is sensitive to all the non-vanishing $\delta_{CP}$ values including 180$^\circ$).
\end{itemize}

The neutrino beam energy spectrum needs therefore to be tuned to measure the oscillatory 
pattern of the flavor conversion process  on the e.g. 1st and 2nd maxima. Referring to Table~\ref{tab:1} we note
that for the shortest baseline CERN-Fr\'ejus, the energy of the 1st maximum is $\simeq 0.26$~GeV. It
grows linearly with distance and reaches $\simeq 4.65$~GeV for the longest baseline
CERN-Pyh\"asalmi. Given the
$L/E$ dependence of the flavor oscillation, the neutrino beam energy should scale with the chosen
baseline $L$ in order to cover those 1st and 2nd maxima.  

\begin{figure} [htb]
\begin{center}
\includegraphics[width=0.99\textwidth]{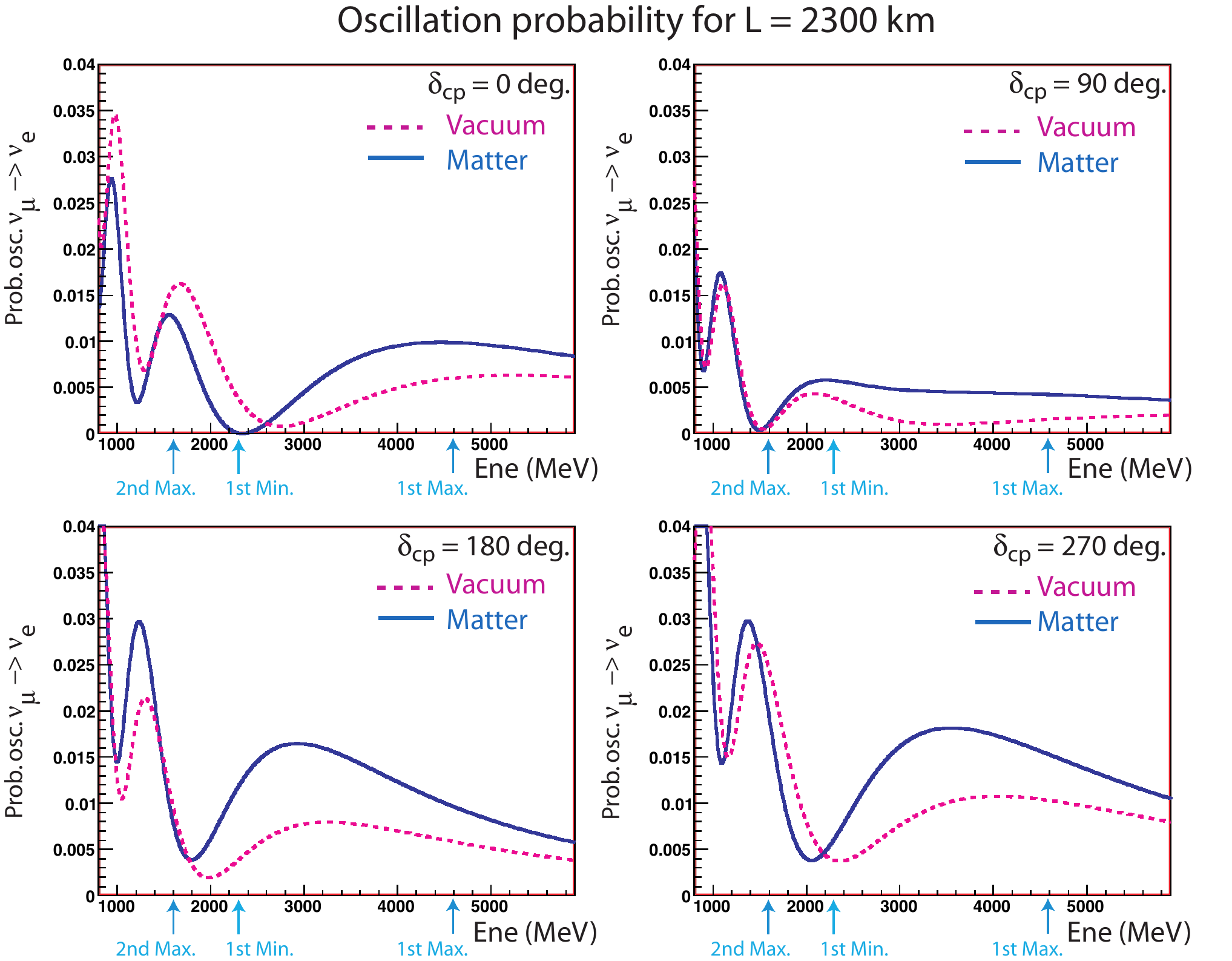} 
\caption{\small Probability of $\nu_{\mu} \to \nu_e$ oscillation for different values of $\delta_{cp}$ without and with matter effects
for $\Delta m^2_{32}>0$ (NH). In this
example, the CERN-Pyh\"asalmi baseline and $\sin^22\theta_{13}=0.01$ were chosen.}
\label{fig:osc2}
\end{center}
\end{figure}

As example, the probability of $\nu_{\mu} \to \nu_e$ oscillation for $\sin^22\theta_{13}=0.01$
and different values of $\delta_{CP}$ with and without matter effects is shown in 
Figure~\ref{fig:osc2} for the CERN-Pyh\"asalmi baseline. The plots 
illustrates qualitatively the fact that a measurement of the oscillation
probability as a function of energy
provides direct information on the $\delta_{CP}$-phase, since this latter introduces
a well-defined energy dependence of the oscillation probability, which is different
from the, say, energy dependence introduced by $\theta_{13}$ alone (when $\delta=0$).
If the neutrino energy spectrum of the oscillated events is experimentally reconstructed with sufficiently 
good resolution in order to distinguish first and second maximum, useful information to extract the 
CP phase is obtained.

We note that the chance to measure both 1st and 2nd maxima increases with the baseline.
Below few hundred MeV (e.g. $\lesssim 400$~MeV) the vanishing cross-section and nuclear
effects including Fermi motion limit the statistics and the energy resolution (for reconstructed 
neutrino energy).  Hence we favor baselines in which both first and second maxima are
above 400~MeV which implies $L\gtrsim 600$~km and a first maximum above $E \simeq 1$~GeV.


\section{Prospects for high intensity neutrino beams from CERN}

To first order of optimization, neutrino rates in conventional beams are proportional to the incident primary proton beam
power~\cite{Ferrari:2002yj}, hence intense neutrino beams can be obtained by trading proton beam
intensity with proton energy. So two basic approaches may be considered: a relatively low proton
energy accompanied by high proton intensity or the second choice is higher proton energy with lower
beam current. The problem at low energy is to obtain the high currents while the challenge
at high energy is the acceleration power to quickly reach the highest energy while maintaining a short 
acceleration cycle.

To second order of optimization the primary proton energy can be tuned to produce the maximum
yield of secondary pions relevant for the production of neutrinos of energy covering the first and
second maximum of the oscillation.

We discuss these points with concrete examples.

\subsection{Upgraded CNGS}
An intensity upgraded CNGS with a low energy focusing optimization~\cite{Rubbia:2002rb} or a coupling to a large detector 
located at an appropriately chosen off-axis position would give
improvements in $\theta_{13}$ reach~\cite{Meregaglia:2006du,Baibussinov:2007ea} compared
to the current optimization for OPERA and ICARUS-T600,  and sensitive searches for CP-violation and neutrino mass hierarchy
determination could be possible~\cite{Meregaglia:2006du}.
These discussions rely on the observation that the CERN SPS has fewer protons and a slower 
cycle than JPARC or FNAL but could accelerate protons  up to 400~GeV with a cycle of 6~s.
Hence, the average beam powers on target are comparable.

An analysis of the maximum potential proton flux from the SPS including possible
upgrade scenarios was consequently performed in Ref.~\cite{Meddahi:2007ju}. We summarize those
results below based on the following assumptions: 200~days per year of operation, 80\% global
machine efficiency and 85\% sharing of the beam (i.e. LHC+neutrinos only). See Table~\ref{tab:accelerators}:
\begin{itemize}
\item Based on operational
experience, it can be estimated that dedicating the SPS FT time to neutrinos
would yield a beam power of 500~kW and $9.4\times 10^{19}$~pots/yr (factor 2 compared to the ``nominal
CNGS").
\item With a double batch injection in the PS from the PSB, one could envisage 
an acceleration of $7\times 10^{13}$ protons in the SPS with a cycle of 7.2~s and
a beam power of 600~kW. This
would yield $11.4\times 10^{19}$~pots/yr (factor 2.5 compared to the ``nominal
CNGS").
\item With the planned new LHC injection chain (see Section~\ref{sec:injupgrade}),
one could envisage accelerating $10^{14}$ protons per SPS cycle. With a new SPS
RF system these protons would be accelerated to 400~GeV every 4.8~s. This
would yield a beam power of 1.3~MW and $24.5\times 10^{19}$~pots/yr (factor 5 compared to the ``nominal
CNGS").
\end{itemize}
Beam losses and equipment heating in the various accelerators and beam lines
will have to be controlled with careful machine tuning and improved controls.

These scenarios are very promising for the future in view of the increased proton
fluxes able to be accelerated in the SPS. However, the bottleneck is 
the intensity limitation of the CNGS infrastructure which without action is essentially limited
to $4.5\times 10^{19}$~pot/yr~\cite{Meddahi:2007ju}. To increase the beam
power of CNGS will require a radiation protection re-classification and/or partial reconstruction 
of its beam-line infrastructure,
raising questions of feasibility,  timescale and costs.
At this stage, the solution to upgrade the CNGS intensity beyond a factor $\times 2$
seems disfavored. Unfortunately it seems 
that potential upgrade scenarios were neglected during the design
of the CNGS facility.

\begin{table}[tb]
\caption{Expected pot per year {[}1e19{]} for different machine scenarios.
$E_{tot}\equiv E_{p}\times N_{pot}$ corresponds to the total amount of energy deposited on the target
per year, which is a relevant quantity to estimate neutrino event rates. }
\label{tab:accelerators} 
\centering{}\begin{tabular}{|l|c|c|c|c|c|c|}
\hline 
 &  PS+SPS & SpS RF & SPL+PS2+ & SPL & New  & Booster +\tabularnewline
 &                 &          upgrade             & SPS new RF & + PS2 & HP-PS &  RCS 4 MW\tabularnewline
\hline
Machine param. & \multicolumn{3}{c|}{\cite{Meddahi:2007ju}} & \cite{ps2wg} & this paper & \cite{Autin2000} \tabularnewline
Proton energy $E_{p}$  & \multicolumn{3}{c|}{400 GeV} & \multicolumn{2}{c|}{50 GeV} & 30 GeV\tabularnewline
\hline
$ppp(\times10^{13})$  & 4.8  & 7 & 10  & 12.5  & 25  &  10\tabularnewline
$T_{c}$ (s)  & 6  & 7.2  & 4.8 & 2.4  & 1.2  &  $(8.33\mathrm{Hz})^{-1}$\tabularnewline
Beam power (MW)  & 0.5  & 0.6  & 1.3  & 0.4  & 1.6  &  4\tabularnewline
Global efficiency  & 0.8  & 0.8 & 0.8 & 1.0  & 1.0  & 1.0\tabularnewline
Beam sharing & 0.85  & 0.85  & 0.85  & 0.85  & 0.85  & 1.0\tabularnewline
Running (d/y)  & 200 & 200  & 200  & 200  & 200   & 200\tabularnewline
$N_{pot}/yr$ (${\bf \times10^{19}}$)  & \textbf{9.4} & \textbf{11.4} & \textbf{24.5} & \textbf{77} & \textbf{300}  & \textbf{1437}\tabularnewline
$E_{tot}\equiv E_{p}\times N_{pot}$  & & & & &  & \\
($\times10^{22}$ GeV$\cdot$pot/yr) & 4  &  4.5 & 10 & 4  & 15   & 43\tabularnewline   
$E_{tot}$ increase  && &   & & & \tabularnewline
compared to CNGS & $\times2$  & $\times2$  & $\times4$  & $\times2$  & $\times5$ & $\times16$ \\
\hline
\end{tabular}
\end{table}

\subsection{Plans for the SLHC injection line -- LP-SPL + PS2}\label{sec:injupgrade}
During the period 2008-2011, a new 160 MeV H- linac (Linac4) will be built to replace the present 50~MeV proton linac (Linac2).
This is the first phase of a plan to renew the LHC injector complex and significantly improve its characteristics~\cite{garoby2007}.
In a second phase, it was proposed to replace
the present 26 GeV PS and its set of injectors (Linac2 + PSB) by
a $\sim 4$~GeV superconducting proton linac (SPL) followed by a $\sim 50$~GeV 
synchrotron (PS2). The SPS itself will be upgraded for 
injection at 50 GeV and for better performance with high brightness beams. 

Beyond the advantages for the LHC luminosity, an order of magnitude higher proton flux at 50 and 4 GeV and 
a new range of possibilities will be available for other users. 
The current SPS-based CNGS programme could significantly profit, as discussed in the previous section,
but the maximum permissible beam intensity onto the CNGS target is limited by the design of its infrastructure
and related radiation safety issues.

With the LP-SPL+PS2 parameters and the upgrade in the SPS intensity, we can conclude the following:
\begin{itemize}
\item The LP-SPL+PS2 and a new SPS RF system could accelerate protons to 400~GeV with a cycle of 4.8~s. This
would yield a beam power of 1.3~MW and $24.5\times 10^{19}$~pots/yr (factor 4 compared to the ``nominal
CNGS").
\end{itemize}

In conclusion, the LP-SPL and PS2 coupled to a new neutrino target and focusing region designed for high power followed by a new decay tunnel  directed towards a LAGUNA site (in the following tentatively called CNXX) could be used to fully exploit the 400~GeV protons 1.3~MW power from the SPS.

\subsection{A high power PS (=HP-PS) or actually HP-PS2 ?}
As discussed previously, the required neutrino beam energy should cover the range $[0,7]$~GeV to
explore the oscillatory behavior of the flavor conversion in the various baseline configurations.
Given the decay kinematics, the relevant parent pions have an energy in the range $[0,15]$~GeV.
This does not require 400~GeV protons and an energy of 50~GeV is sufficient to kinematically produce mesons of the 
relevant energies.

Actually 50~GeV protons produce a neutrino spectrum with less tail at high energy
than 400~GeV protons. This is an advantage when considering backgrounds from neutral current interactions
in the far detectors.

We can hence argue that:
\begin{itemize}
\item  The baseline FT parameters defined by the PS2 working group~\cite{ps2wg}
yield $1.2\times 10^{14}$ protons at 50 GeV with a cycle
of 2.4~s. This would correspond to a beam power of 0.4~MW and $7.7\times 10^{20}$~pots/yr.
In terms of $E_{tot}\equiv E_p\times N_{pot}$ this gives a factor of 2 compared to the ``nominal
CNGS". There is therefore almost no gain compared to the optimized PS+SPS scenario, although
the benefit of relieving the present SPS from demanding high intensities could be an advantageous choice
for the long term operation for the LHC.
\item A high power PS2 (=HP-PS2), whereby a factor 4 in intensity compared to the baseline parameters is assumed,
could be achieved by doubling the proton intensity  and doubling the repetition rate. This would yield a beam power of 1.6~MW and $3\times 10^{21}$~pots/yr. In terms of $E_{tot}\equiv E_p\times N_{pot}$ this corresponds to a factor 5 compared to the ``nominal
CNGS".
\end{itemize}

The possibility to change the design of the PS2 to power upgrade necessitates a dedicated critical study: is it necessary to
consider an increase of the machine aperture (with impact on magnets, etc.)?
Could a higher injection energy be envisaged to reduce space charge at injection? 
A certain fraction of the desired intensity increase could be obtained by ``operational experience'': which fraction ?
The increase of the repetition rate would imply an upgrade of the magnet power supplies
and of the RF system, which could be implemented later. Is there sufficient space reserved for the
accelerating regions? All these questions need careful answers but ultimately there does not seem
to be any technical show-stopper. Operational experience, e.g. at the J-PARC MR, will in the coming years
certainly provide very valuable information.

\subsection{High energy rapid cycling synchrotrons ?}
The LP-SPL+PS2 design was proposed as presenting significant advantages in the CERN context, 
especially because of its flexibility and its capability to evolve towards the very large beam power.
On the other hand, rapid cycling synchrotrons (RCS) are being developed as an alternative
solution to reach high power at high energy and were featured in proposals from Brookhaven and
RAL (see e.g.~\cite{Prior:2006p1187}). 
JPARC is presently commissioning a 25~Hz 3~GeV RCS as injector to the 30~GeV Main Ring~\cite{Hasegawa:2010xy}.
There is no doubt that a 8~Hz  30~GeV proton RCS with
a power of 4~MW as discussed previously in the CERN context~\cite{Autin2000} would provide ``ultimate'' superbeam performance for
long baseline experiments and could be used as the proton driver of the neutrino factory. 
This might well be the correct path for high power at CERN. 
However, in the following, we will focus on the physics performance of a 1.6~MW HP-PS2.
A 4~MW proton source would reduce the running time of the experiment accordingly.


\section{Expected neutrino oscillation physics performance of a European long baseline experiment based on a CERN 1.6~MW HP-PS2 superbeam}\label{sec:expphys}

Based on the previous arguments, we now discuss the expected physics performance of long
baseline neutrino physics based on a 1.6~MW HP-PS2 based at CERN. Our focus is on $\theta_{13}$
determination, CP-violation discovery and neutrino mass hierarchy determination.

\begin{figure}[htbp]
\centering
\includegraphics[width=0.95\textwidth]{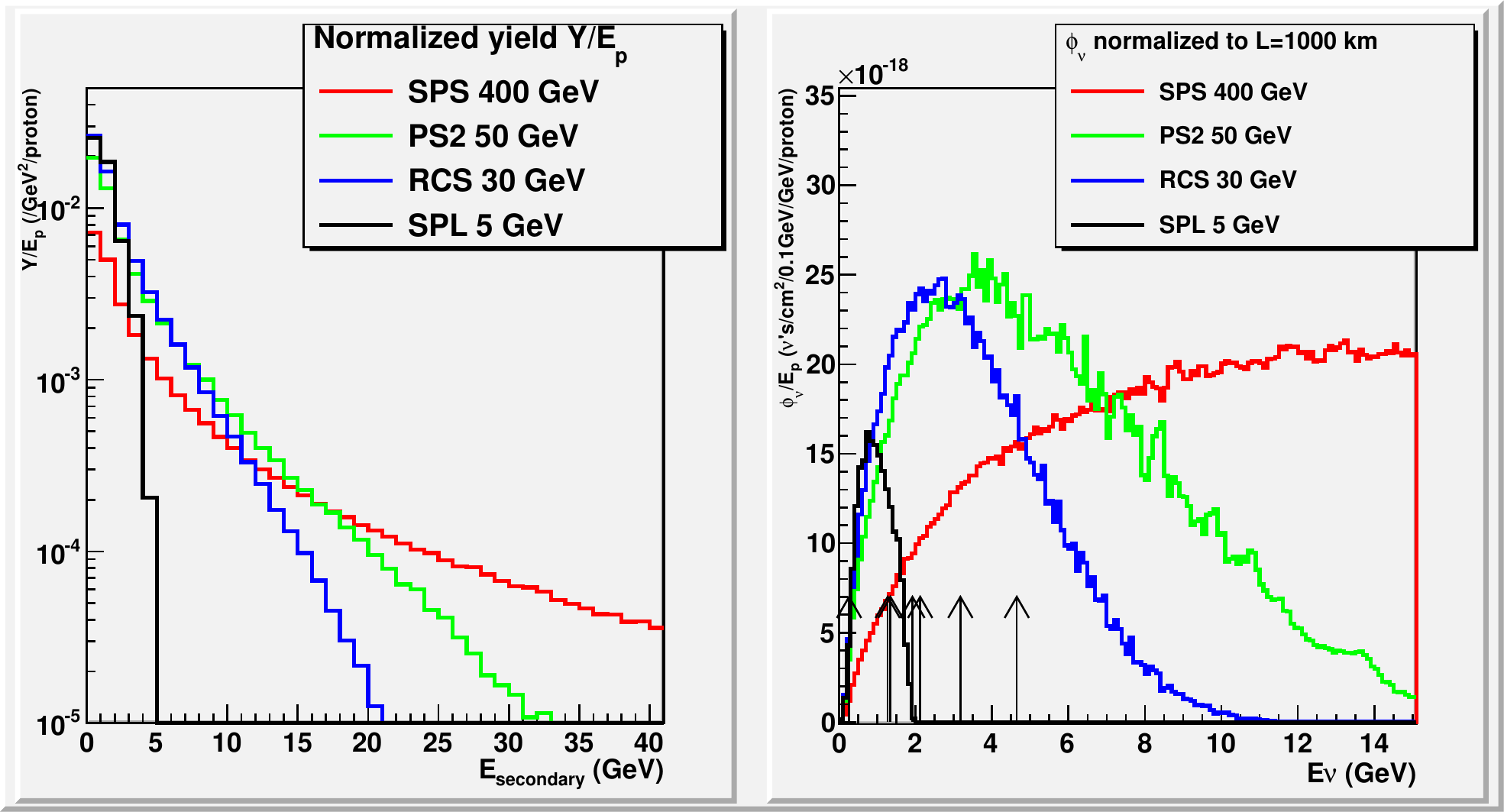}
\caption{(left) Normalized pion production rate $Y/E_p$ for various
incident proton energies $E_p$ as estimated with GEANT4; (right)
Normalized neutrino flux $\phi_{\nu}/E_p$ arbitrarily normalized to a baseline of 1000~km.
The vertical arrows indicate the energy of the 1st maxima of the 
neutrino oscillations for the 7 different baselines considered in LAGUNA
(see text).}
\label{fig:piyield}
\end{figure}

In order to have a preliminary quantified assessment of the physics performance, 
a GEANT4-based~\cite{g4} simulation was developed in order to compare the meson yields in proton interactions
of 5, 30, 50 and 400~GeV incident energy. A graphite target was chosen since it is widely used in current
experiments. It has a density of $\rho=2.2$~g/cm$^3$, a diameter of 4~mm and
a length of 100~cm. The power dissipation in the target was not addressed. The standard
GEANT4 reference physics model (QGSP\_BERT 3.3) has been used, in absence of a better
choice. Secondary mesons produced in the interactions crossing a 1~$m^2$ area behind
the end of the target were recorded during the simulation and used for computation
of the neutrino flux.
In order to compare the proton energy options, the secondary meson production yield was normalized by 
the incident proton energy $E_p$. The resulting yields $Y/E_p$ in particles/GeV$^2$/proton
are shown in Figure~\ref{fig:piyield}(left). For secondary meson energies below 5~GeV the 
energy scaling law is rather remarkable since the 5, 30 and 50~GeV curves
almost overlap. At higher energies the tail increases as the 
incident proton energy increases. At 400~GeV the tail extends very high above the
relevant secondary pion range defined as [0,15]~GeV. It is also evident that
400~GeV produce less low energy secondaries per proton than at lower
proton energies. In this sense, the SPS energy is not optimized as it produces
several very high energy pions. We will attempt to recover this situation by considering
an off-axis position.

We initially computed the neutrino fluxes assuming ideal focusing.
In this case, all secondaries reaching the 1~$m^2$ area behind
the target were ideally focused and then decayed
to produce neutrinos. Ideal focusing implies
that their 3-momentum was rotated around the meson position in the transverse
plane till their transverse momentum vanishes and the decay path was ignored. 
The ideal focusing calculation is an important method to optimize the proton incident energy by maximizing 
the yield of secondaries in the relevant energy. One can subsequently design
a focusing system to best match the ideal conditions. 
In our simulation, the neutrinos crossing a 100~$m^2$ area detector placed at an arbitrary
distance of 1000~km is used to compute the resulting flux. The flux are also normalized
to the incident proton energy in order to easily compare the 5, 30, 50 and 400~GeV incident
proton curves. These normalized neutrino fluxes $\phi_{\nu}/E_p$ are plotted
in Figure~\ref{fig:piyield}(right).  The vertical arrows indicate the energy of the 1st maxima of the 
neutrino oscillations for the 7 different baselines considered in LAGUNA. 

We conclude from these plots that the 30 and 50~GeV incident proton energies are the most
adequate to produce a wide band neutrino beam neutrino with the desired energy coverage.

\begin{figure}[htbp]
\centering
\includegraphics[width=0.95\textwidth]{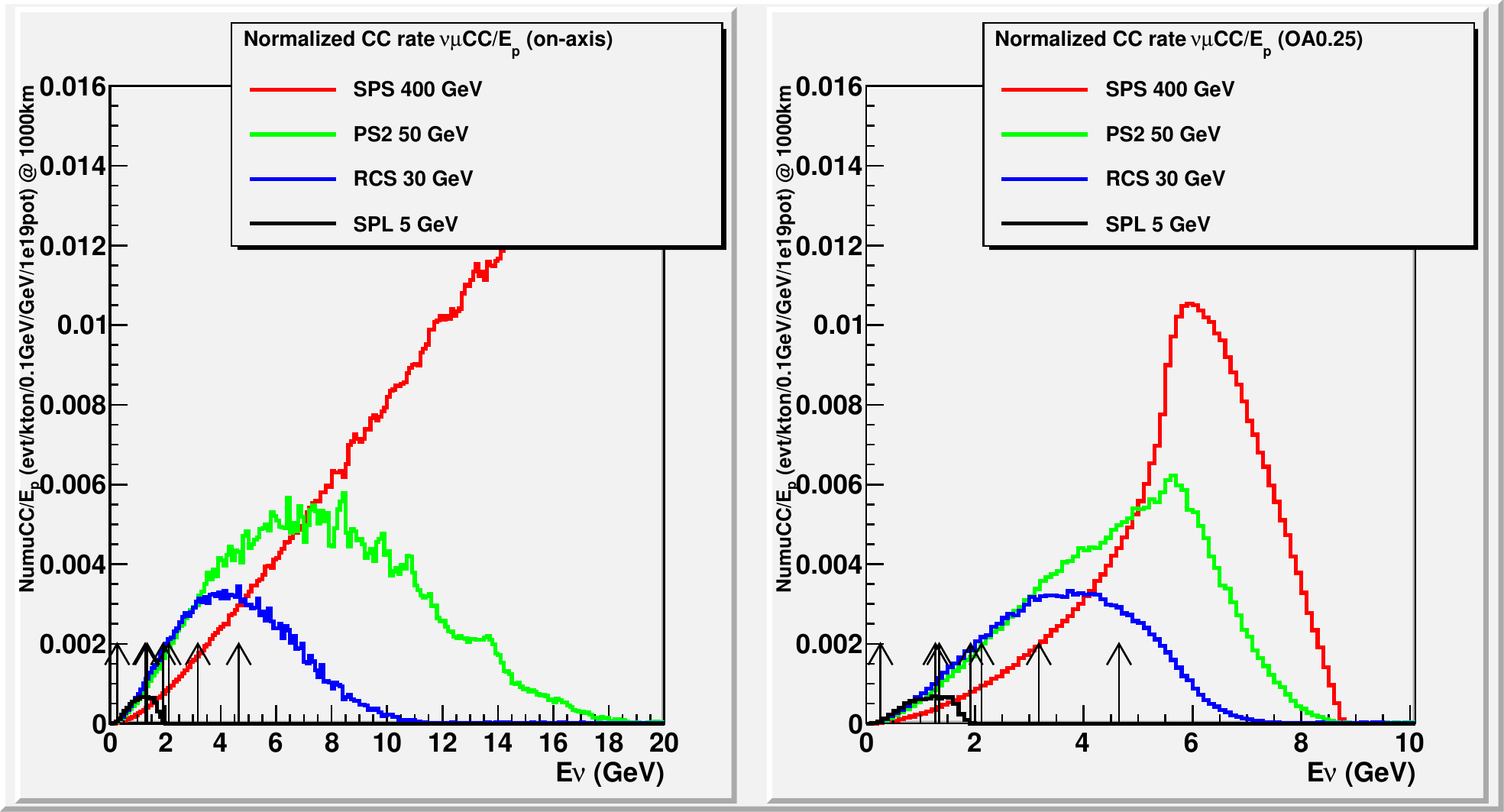}
\caption{(left) Normalized energy spectrum of beam muon neutrino charged current interactions 
in absence of neutrino oscillations $\nu_\mu$~CC/$E_p$;
(right) same for an off-axis 0.25$^o$ configuration.
}
\label{fig:eventdistrideal}
\end{figure}

We then use the NUX\footnote{The NUX neutrino-nucleon interaction code 
was developed to produce neutrino-nucleus interactions, including quasi-elastic, resonance 
production and deep inelastic scattering. The combined FLUKA(PEANUT)+NUX model gave 
outstanding results when compared with NOMAD data.} cross-sections in order to compute 
the  $\nu_\mu$ CC event rate and their energy distribution (in absence of neutrino
oscillations). The results are shown in Figure~\ref{fig:eventdistrideal}(left).
The 30 and 50~GeV protons produce the best results, while the beam produced by 
400~GeV protons is too hard. 

In order to compensate for this problem, we can consider the ``off-axis'' technique, 
pioneered in the Brookhaven neutrino oscillation experiment proposal~\cite{BNL}
and used in T2K and NOvA, which
consists of placing a neutrino detector at some angle with respect to the
conventional neutrino beam. An ``off-axis'' detector records approximately the same 
flux of low energy neutrinos, as the one positioned ``on-axis'', originating from the decays 
of low energy mesons. In addition, though, an ``off-axis'' detector 
records an additional contribution
of low energy neutrinos from the decay of higher energy parents decaying at a
finite angle. 

The effect of an 0.25$^o$ off-axis configuration is illustrated 
in Figure~\ref{fig:eventdistrideal}(right), where the vertical axis
has the same scale as for the on-axis configuration (same Figure (left)).
We can see that a small off-axis angle is very effective at suppressing
high-energy neutrinos. This method maintains the flux relatively unchanged
and a broad neutrino spectrum is obtained, quite convenient for optimally 
measuring the first and second oscillation maxima. 

Similar conclusions as for the on-axis apply here although the off-axis
angle is effective at rescuing the 400~GeV SPS case. This can
be a good solution to allow the same new beam-line to operate
initially with the (existing) SPS until the HP-PS2 is commissioned.
Indeed, the present foreseen localization of the PS2 ring,
its extractions lines and the possible associated experimental
areas~\cite{Baldy2007}  should make it conceivable to extract via appropriate
transfer lines both PS2 and SPS beams into the new CNXX beam target area, in particular
if the target and the decay tunnel are oriented towards
a north-European far location.

\section{Physics performance with a realistic horn focusing}

In order to assess the physics performance based on the HP-PS2 with a realistic -- although
not fully optimized geometry -- horn focusing system, we implemented a
magnetic focusing system in our simulation, with similar geometry as that of the NUMI beamline~\cite{Abramov:2001nr},
choosing as a starting point the NUMI-ME configuration. At this stage, we 
concentrate on three baselines CERN-Sierozsowice (950~km), CERN-Slanic (1544~km)
and CERN-Pyh\"asalmi (2300~km). Results for the other baselines will be reported later.

A fast particle tracking programme which neglects interactions of secondaries in the focusing and
other beam line materials, has been used to rapidly estimate the neutrino fluxes and integrate
them to compute the expected event rates. These latter are summarized in Table~\ref{tab:EventsNoOsc}
in case of no flavor oscillations. The figures are calculated for the NUMI-ME-like realistic focusing, 
  normalized for one year and a liquid Argon detector with a mass of 100~kton.

The CERN-Slanic (1544~km) and CERN-Pyh\"asalmi (2300~km) were envisioned with an off-axis
angle of 0.25$^o$ while the CERN-Sierozsowice (950~km) has twice the off-axis
angle 0.5$^o$ in order to a neutrino beam spectrum of lower energy given the
shorter distance. 

The event rates are in the range of 10'000 to 20'000 neutrino events
per year for a positive horn polarity (neutrino run)
and about 1/3 antineutrino events (for 50~GeV) to about 1/2 antineutrino events (for 400~GeV)
for the opposite polarity of the horn (antineutrino run), assuming the same number of pots
for each polarity.
One a priori advantage of the high energy is the relatively symmetric production of positively and
negatively charged pions. While at 50~GeV, leading charge effects in the fragmentation 
leads to an excess of $\pi^+\rightarrow \nu$'s compared to  $\pi^-\rightarrow \bar\nu$'s.
As a result, the rate of events in the antineutrino run is more suppressed for 50~GeV protons
than for 400~GeV.
However, the contamination of neutrino events in the antineutrino run ($\nu_\mu$ CC contamination
relative to $\bar\nu_\mu$ CC in antineutrino run) is more favorable for 50~GeV protons than
for 400~GeV because very forward meson production is a dominant source of 
wrong helicity neutrinos at high energy.

When computing oscillation sensitivities, we include for each run polarity, the opposite polarity 
particles which are not defocalized are included in the calculation and added to the rate since 
we neglect at this stage the experimental determination of the 
helicity of the incoming neutrino (or the charge of the outgoing lepton).
Detailed oscillation sensitivity calculations (as those detailed below) show that
the opposite neutrino helicity contamination plays a non-negligible role in the 
sensitivity to CP-violation. Hence, from that point of view, we favor 50~GeV protons over 
400~GeV and from now on focus on this case.

\begin{table}[htb]
  \begin{center}
    \begin{tabular}{|c|c|c|c|c|c|c|}
         \hline
         &  \multicolumn{3}{c|}{neutrino run} & \multicolumn{3}{c|}{antineutrino run} \\
         \hline
            Distance/OA	 & $\nu_{\mu}$CC  &	$\nu_e$CC  &	($\nu_{e} + \overline{\nu}_e$) / & $\nu_{\mu}$CC  &	$\nu_e$CC  &	($\nu_{e} + \overline{\nu}_e$) / \\     
       &  ($\overline{\nu}_{\mu}$CC) & ($\overline{\nu}_e$CC) & ($\nu_{\mu} +  \overline{\nu}_{\mu}$) &  ($\overline{\nu}_{\mu}$CC) & ($\overline{\nu}_e$CC) & ($\nu_{\mu} +  \overline{\nu}_{\mu}$)\\
       \hline
          \multicolumn{7}{|c|} {	\bf{CNXX NUMI-ME-like horns}  , 400 GeV SPS protons , 2.4$\times 10^{20}$ pot/year}\\
	 \hline
	          \bf{1544} km & & && & &\\
    0.25 deg   & 12181 (939)& 96 (16) & 0.9 $\%$& 2469 (5125)& 37 (39) & 1.0 $\%$\\
          \hline
    \multicolumn{7}{|c|} {	\bf{CNXX NUMI-ME-like horns}  , 50 GeV HPPS2 protons , 3$\times 10^{21}$ pot/year}\\
	 \hline
    	\bf{950} km &  & && & &\\
           0.5 deg & 22167 (327) & 165 (9) & 0.8 $\%$& 1270 (6068) & 27 (43) & 1.0 $\%$\\
            \hline
            \bf{1544} km &  & && & &\\
           0.25 deg & 23600 (333) & 160 (7) & 0.7 $\%$& 1267 (6467) & 20 (40) & 0.8 $\%$\\
            \hline
    	\bf{2300} km &  & && & &\\
           0.25 deg & 10667 (153) & 73 (3) & 0.7 $\%$& 573 (2933) & 7 (20) & 0.8 $\%$\\
       \hline
      \end{tabular}
  \caption{Event rate calculated with a NUMI-ME-like realistic focusing, 
  normalized for one year and a liquid Argon detector with a mass of 100~kton.}
    \label{tab:EventsNoOsc}
  \end{center}
\end{table}

\begin{figure}[htbp]
\centering
\includegraphics[width=0.495\textwidth]{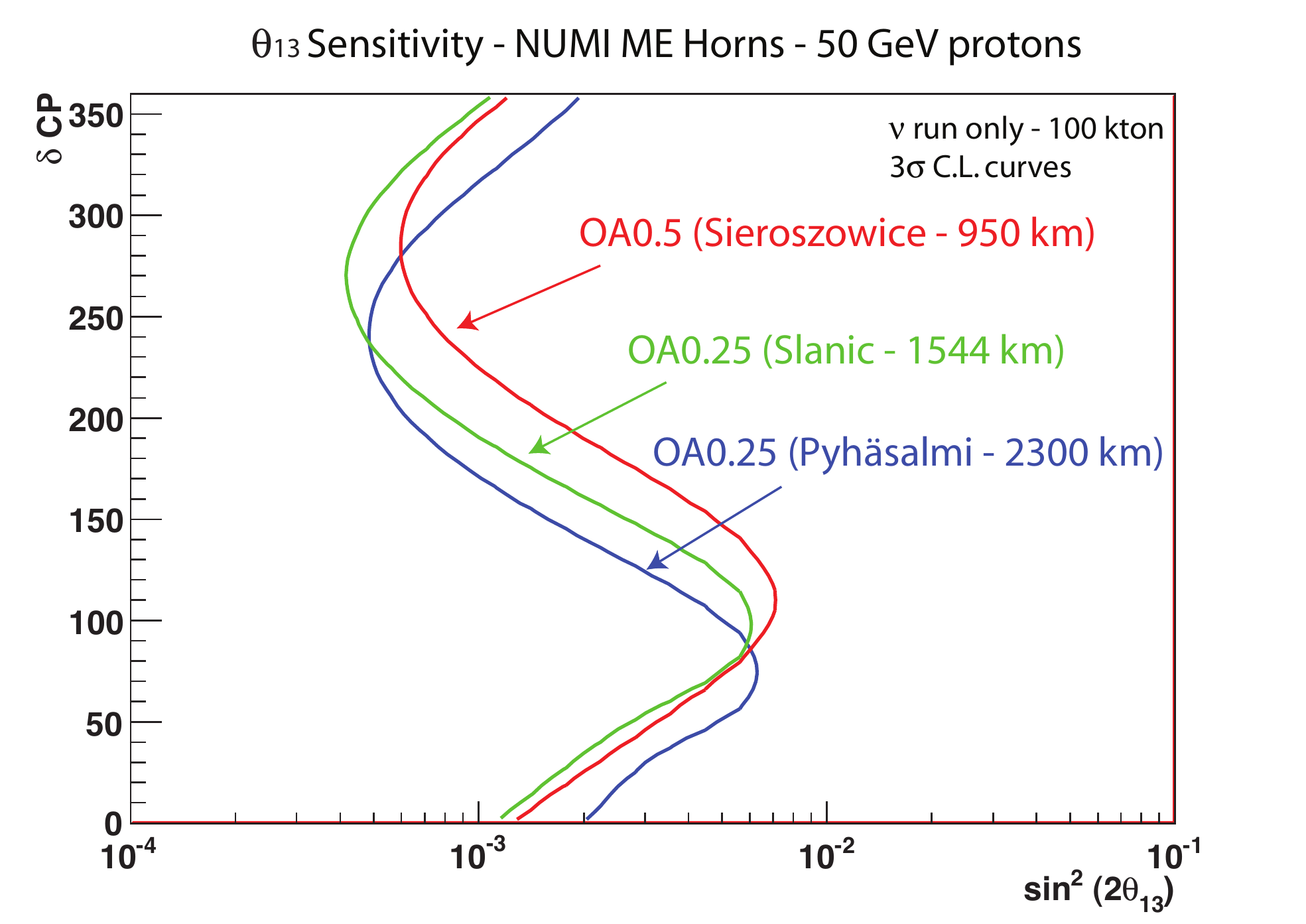}
\includegraphics[width=0.495\textwidth]{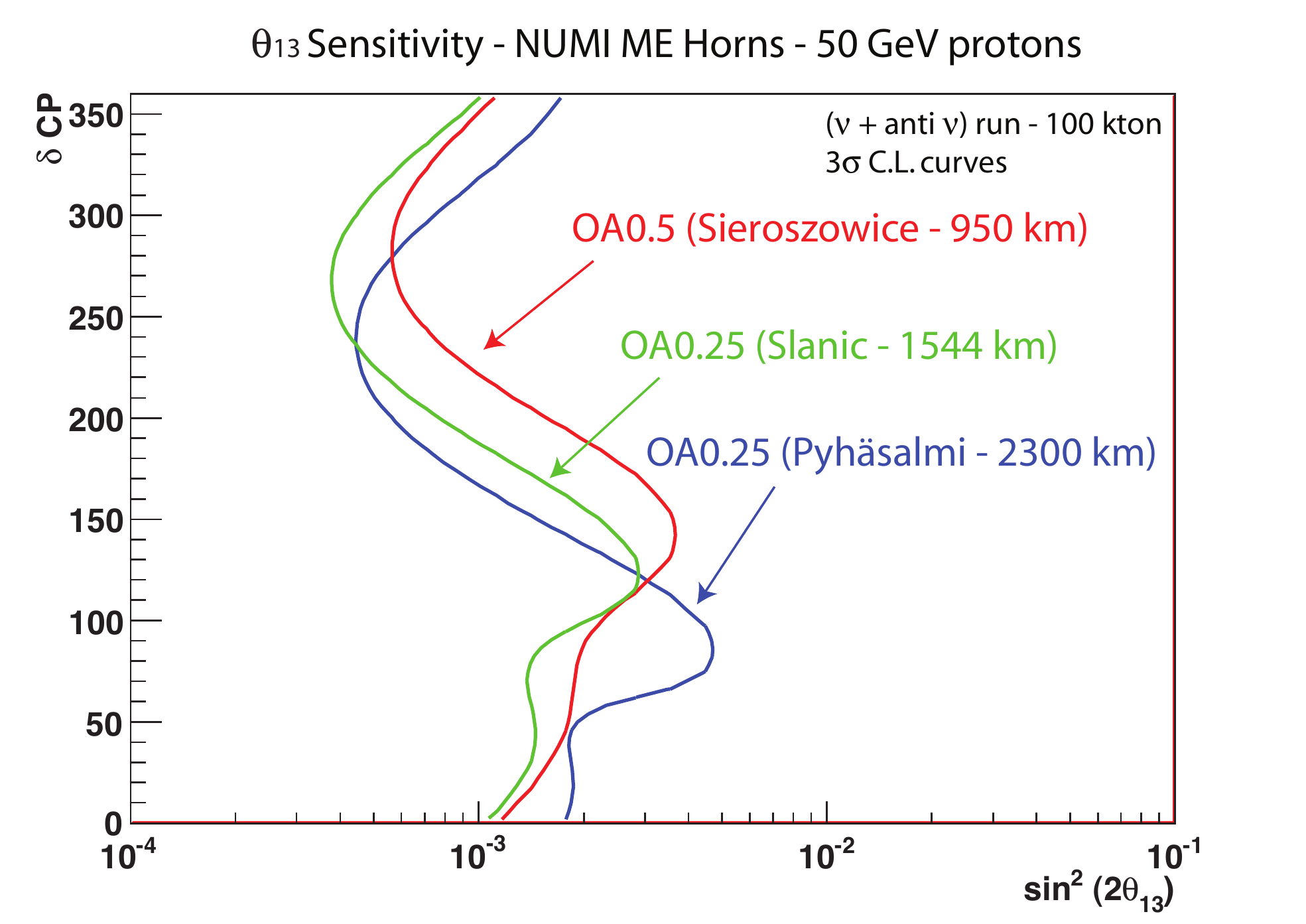}
\caption{(left) Discovery potential at $3\sigma$ for $\theta_{13}$, for CNXX (NUMI-ME-like Horns 50 GeV HP-PS2 protons) beam for three LAGUNA locations, 3 $\times$ 10$^{21}$ pot/year and 5 years of neutrino run; (right) same but with 5 years of neutrino run plus 5 years of anti neutrino run.}
\label{fig:disc_theta}
\end{figure}

\begin{figure}[htbp]
\centering
\includegraphics[width=0.495\textwidth]{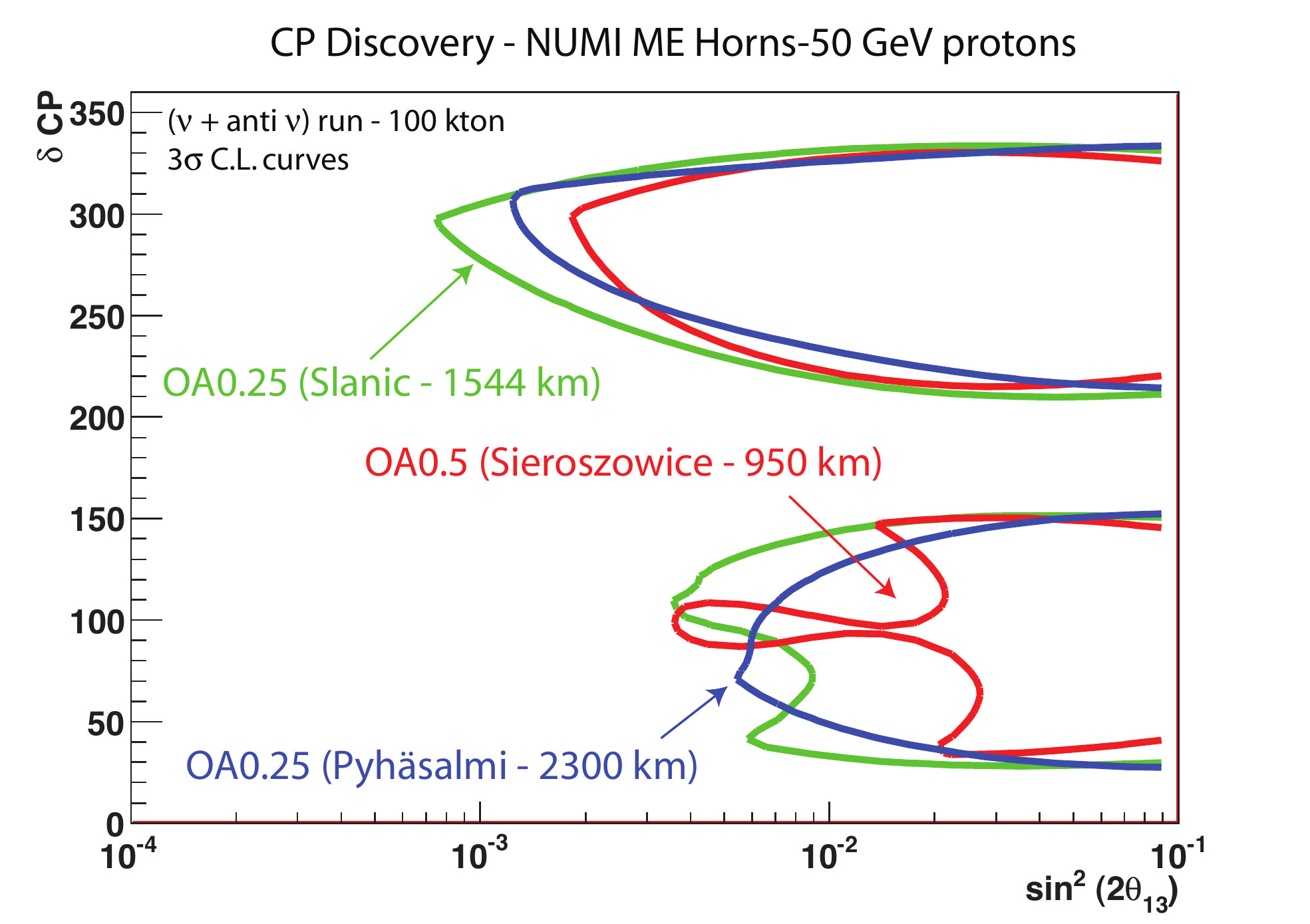}
\includegraphics[width=0.495\textwidth]{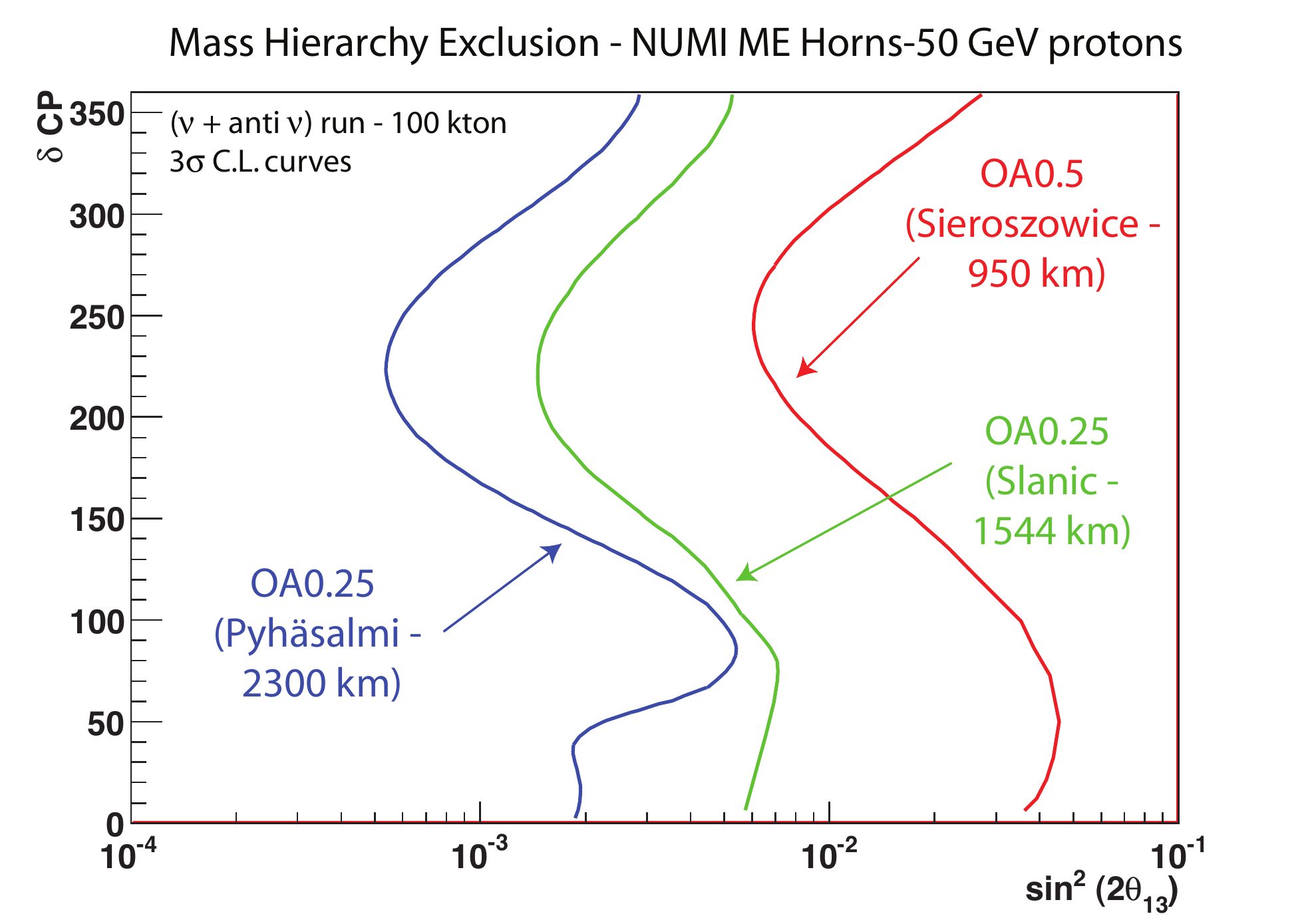}
\caption{(left) Discovery potential at $3\sigma$ for  CP-violation, for CNXX (NUMI-ME-like Horns 50 GeV HP-PS2 protons) beam for three LAGUNA locations, 3 $\times$ 10$^{21}$ pot/year and 5 years of neutrino run plus 5 years of anti-neutrino run; (right) Mass hierarchy discrimination at $3\sigma$
for same beam conditions.}
\label{fig:discCP}
\end{figure}

The expected sensitivities were computed with the help of the GLOBES~\cite{Huber:2004ka} software,
in a similar way to what we did previously which assumed a 100~kton liquid Argon detector
(see Ref.~\cite{Meregaglia:2006du} for details):
\begin{itemize}
\item In order to discover a non-vanishing
$\sin^22\theta_{13}$, the hypothesis $\sin^22\theta_{13}\equiv 0$
must be excluded at the given C.L. As input, a true non-vanishing value
of $\sin^22\theta_{13}$ is chosen in the simulation and a fit with
$\sin^22\theta_{13}= 0$ is performed, yielding the ``discovery''
potential. This procedure is repeated for every point in the ($\sin^22\theta_{13},\delta_{CP}$)
plane.
The corresponding sensitivity to discover $\theta_{13}$ in the true ($\sin^22\theta_{13},\delta_{CP})$
plane at $3\sigma$ is shown in Figure~\ref{fig:disc_theta}. 
The (left) panel shows the sensitivity with 3 $\times$ 10$^{21}$ pot/year and 
5 years of neutrino run; the (right) panel assumes 
5 years of neutrino run plus 5 years of anti neutrino run.
$\Longrightarrow$ The $\theta_{13}$ sensitivity is in first approximation independent of the baseline since
the decrease in flux with increasing squared distance is compensated
by the increased neutrino cross-section with energy.
\item By definition, the CP-violation in the lepton sector can be said to be
discovered if the CP-conserving values, $\delta_{CP}=0$ and $\delta_{CP}=\pi$, 
can be excluded at a given C.L. The reach for discovering CP-violation is computed
choosing a ``true'' value for $\delta_{CP}$ ($\ne 0)$ as input at different true values
of $\sin^22\theta_{13}$ in the $(\sin^22\theta_{13},\delta_{CP})$-plane,
and for each point of the plane calculating the corresponding 
event rates expected in the experiment. This data is then fitted with
the two CP-conserving values $\delta_{CP}=0$ and $\delta_{CP}=\pi$, leaving
all other parameters free (including $\sin^22\theta_{13}$~!).
The opposite mass hierarchy is also fitted and the minimum of all
cases is taken as final $\chi^2$. 
The corresponding sensitivity to discover CP-violation 
in the true ($\sin^22\theta_{13}, \delta_{CP})$
plane is shown in Figure~\ref{fig:discCP}(left). 
At the shorter baseline, matter effects are at the level of 30~\%, hence
it can be difficult to detect and untangle this effect from CP-phase induced asymmetries.
Indeed, for certain combinations of true $\sin^22\theta_{13}$ and $\delta_{CP} $,
it is possible to fit the data with the wrong mass hierarchy and a rotated $\delta_{CP} $ by 90$^o$,
an effect labelled as $\pi$-transit~\cite{Huber:2002mx}.
$\Longrightarrow$ The ability to  discover CP-violation improves with the baseline,
in particular as we approach the ``magical'' distance of Ref.~\cite{Raut:2009jj}
as in the case of the CERN-Pyh\"asalmi baseline.
\item In order to determine the mass hierarchy to a given C.L., the opposite
mass hierarchy must be excluded. A point in parameter space with
normal hierarchy is therefore chosen as true value and the solution
with the smallest $\chi^2$ value with inverted hierarchy has to
be determined by global minimization of the $\chi^2$ function
leaving all oscillation parameters free within their priors. 
The sensitivity to exclude inverted mass hierarchy
in the true ($\sin^22\theta_{13},\delta_{CP})$
plane is shown in Figure~\ref{fig:discCP}(right). $\Longrightarrow$As expected,
the sensitivity to the neutrino mass hierarchy improves with the baseline.
\end{itemize}

\section{Conclusions}
With the eminent startup of neutrino reactor experiments as well as T2K and NOvA, 
one cannot exclude that a positive $\theta_{13}$ signal will be found
in the next few years. This
would indeed mean that $\sin^{2}2\theta_{13}\gtrsim 0.01$, a value compatible
with the interpretation of solar and atmospheric neutrino data~\cite{Fogli:2008jx}, 
the recent MINOS appearance analysis~\cite{MINOS:2009yc} and
the latest SNO result~\cite{SNO:2009gd}.
A positive signal would certainly provide a tremendous boost to 
long-baseline neutrino physics, opening the prospects to detect
CP-violation with superbeams. In case that no evidence for  $\theta_{13}$ is found
in the current round of experiments, it will be worth to
continue the quest with more intense superbeams (=``hyperbeams'') than presently envisaged
for T2K and NOvA, and with more massive
far neutrino detectors, allowing to improve the $\theta_{13}$ sensitivity and proton lifetime
by at least an order magnitude.

Neutrino factories and beta-beams certainly represent fantastic precision machines that the neutrino
community should hope for, however until their R\&D is successfully accomplished,
conventional neutrino superbeams remain technically proven solutions
which can be accurately costed and reliably implemented on a timescale of
a decade, even in the multi-MW regime, offering concrete plans to move
forward.

From the different available options for superbeams, high-energy and wide-band beams
associated to the longest baselines provide the best opportunities to reach the
experimental goals, in some cases avoiding the necessity of antineutrino runs. 
The optimal proton source energy lies in the range $30\div 50$~GeV.

Both KEK/JPARC and FNAL are considering upgrade paths for their ongoing
neutrino programmes along the lines described here. The European programme
could become attractive with the very long baselines not elsewhere accessible ($\gtrsim 1300$~km), and 
because of the flexibility offered by designing a completely new beamline (target,
optics, off-axis angle, etc.), 
naturally provided that the timescale does not exceed the $\sim 2025$ horizon.

We therefore believe that a CERN HP-PS2 proton source should be
envisaged since it would represent the most time- and cost-effective way
to establish forefront long baseline neutrino physics in Europe with prospects
of discovering CP-violation in the leptonic sector.

\section*{Acknowledgements}
The author acknowledges the support from the JSPS Invitation Training Program for 
Advanced Japanese Research Institutes.
The LAGUNA design study is financed by FP7 Research Infrastructure "Design Studies", Grant Agreement No.
212343 FP7-INFRA-2007-1.
The author also benefitted from useful discussions with R.~Garoby (CERN) and M.~Benedikt (CERN), and 
from the help of A.~Meregaglia for the GLOBES calculations.

\end{document}